# High-power pulsed electrochemiluminescence for optogenetic manipulation of *Drosophila* larval behaviour

Chang-Ki Moon[1,2], Matthias König[1,2], Ranjini Sircar[1,3], Julian F. Butscher[1,2], Stefan R. Pulver[3], Malte C. Gather[1,2]

[1]Humboldt Centre for Nano- and Biophotonics, Institute for Light and Matter, Department of Chemistry, University of Cologne, Greinstr. 4-6, 50939 Cologne, Germany
[2]Centre of Biophotonics, SUPA, School of Physics and Astronomy, University of St Andrews, North Haugh, St Andrews KY16 9SS, United Kingdom
[3]School of Psychology and Neuroscience, University of St Andrews, St Mary's St Mary's Quad, South St, St Andrews KY16 9JP, United Kingdom

Electrochemiluminescence (ECL) produces light through electrochemical reactions and has shown promise for various analytic applications in biomedicine. However, the use of ECL devices (ECLDs) as light sources has been limited due to insufficient light output and low operational stability. In this study, we present a high-power pulsed operation strategy for ECLDs to address these limitations and demonstrate their effectiveness in optogenetic manipulation. By applying a biphasic voltage sequence with short opposing phases, we achieve intense and efficient ECL through an exciplex-formation reaction pathway. This approach results in an exceptionally high optical power density, exceeding 100 µW/mm², for several thousand pulses. Balancing the ion concentration by optimizing the voltage waveform further improves device stability. By incorporating multiple optimized pulses into a burst signal separated by short rest periods, extended light pulses of high brightness and with minimal power loss over time were obtained. These strategies were leveraged to elicit a robust optogenetic response in fruit fly (*Drosophila melanogaster*) larvae expressing the optogenetic effector CsChrimson. The semi-transparent nature of ECLDs facilitates simultaneous imaging of larval behaviour from underneath, through the device. These findings highlight the potential of ECLDs as versatile optical tools in biomedical and neurophotonics research.

# Introduction

Organic semiconductors offer versatile luminescent properties and organic light-emitting diodes (OLEDs) based on these have found widespread applications in lighting, mobile displays, augmented/virtual reality, and biophotonics[1-3]. Electrochemiluminescence (ECL) is a related branch of the organic semiconductor technology, leveraging molecular luminescence through redox reactions in liquids or gels[4,5]. ECL has been applied across various fields of biomedical research for sensing, diagnostics and in various analytical methods[6-10]. In addition, ECL devices (ECLDs) have the potential to serve as flexible light sources for other areas of biomedicine—such as optogenetics—that depend on intense and sustained illumination to either monitor intracellular processes or control the activity of excitable cells. Such applications often require tailored device designs for effective light delivery to both superficial and deep tissues[11-15].

ECLDs can be fabricated via a liquid injection technique that offers several advantages over conventional vacuum evaporation and solution processing methods used for the fabrication of solid-state light-emitting devices like OLEDs. These include a simplified manufacturing process, cost-effectiveness, minimal waste generation, and adaptable design configurations[16-19]. However, despite these benefits, ECLDs have historically received little attention as light sources due to their relatively low light output and limited operational stability.

High-power ECLDs require an abundant production of radical ions and a rapid reaction between these ions. Unfortunately, high-voltage and/or prolonged operation of ECLDs often leads to an accumulation of radical ions in the device, thus triggering side reactions that limit device performance and lifetime[20]. To mitigate the accumulation of ions, various strategies have been explored; these include operating devices under alternating current (AC) to accelerate interionic annihilation[21,22], tuning AC waveforms to maintain balanced concentrations of redox species[23], and inserting a rest period between DC voltage sequences to prevent build-up of radical ions[24,25]. Recently, we demonstrated a substantial improvement in ECLD performance through the introduction of an exciplex-formation pathway[26], in which exciplex materials at high concentrations mediate redox reactions and recombination while the light emission itself is achieved by energy transfer to a dye molecule present at a relatively low concentration. This approach significantly improved device stability compared to the conventional interionic annihilation pathway; under AC operation, these exciplex devices achieved a luminance of up to 1250 cd/m$^2$ and an optical power density (OPD) of up to 9.1 µW/mm$^2$. Despite these advances, applying a higher voltage, e.g. above 4 V, remains challenging in typical sandwich electrode devices, as this can lead to side reactions occurring at a significant rate, which can cause material degeneration and prevent reaching sufficiently high optical output for some applications in biomedical research. Moreover, the operational lifetime of ECLDs has so far been limited to a few minutes, even at a low brightness level of 100 cd/m$^2$, with a further severe decline in lifetime at higher brightness.

Here, we present high-voltage and high-power pulsed operation of ECLDs and demonstrate their utility in optogenetic manipulation of the behaviour of *Drosophila* larvae. Applying a biphasic voltage sequence to an ECLD based on exciplex materials induces rapid and efficient ECL near the electrode surfaces, resulting in intense light emission. For ±10 V pulses of 0.2 ms duration per phase, we achieve a high OPD exceeding 100 µW/mm$^2$ over several thousand pulses. The strong improvement in performance achieved upon optimizing the voltage waveform highlights the importance of balancing ion concentrations to maximize operational longevity. Furthermore, incorporating multiple ECL pulses into a burst signal, with short rest periods between voltage sequences, extends the duration of emission to several seconds with minimal optical power loss. Using this approach, an ECL pulse train, configured to provide quasi-continuous illumination for four seconds, elicited robust body-bending responses in *Drosophila* larvae expressing the light sensitive ion channel CsChrimson in a set of interneurons which trigger escape behaviours. By placing the larvae directly on an ECLD pixel and observing from underneath using infrared light passing through the ECLD, we can simultaneously observe the bending and twisting of larvae and deliver light pulses with high temporal and spatial resolution.

## Results

### Pulsed electrochemiluminescence

To ensure our ECLDs reach high optical transmittance as required for imaging and microscopy through the device, we fabricated ECLDs in a configuration of glass(0.7 mm)/ITO(100 nm)/ECL solution(30 μm)/ITO(100 nm)/glass(0.7 mm). The ECL solution consists of 30 mM 1,1-bis[(di-4-tolylamino)phenyl]cyclohexane (TAPC, donor), 30 mM 2,2',2''-(1,3,5-benzinetriyl)-tris(1-phenyl-1-H-benzimidazole) (TPBi, acceptor), 10 mM 2,8-di-tert-butyl-5,11-bis(4-tert-butylphenyl)-6,12-diphenyltetracen (TBRb, yellow-emissive dye), and 100 mM tetrabutylammonium hexafluorophosphate (supporting electrolyte), all dissolved in a 2:1 by volume mixture of toluene and acetonitrile. This device configuration yielded high optical transmittance (~80%) in the region ranging from 580 nm to 1000 nm as shown in Fig. 1a. The absorption peaks at 491 nm and 525 nm correspond to the absorption of TBRb and reduce transmittance to approximately 40% in this spectral band. The photograph in the inset shows a logo of our research centre as seen through an ECLD placed under normal illumination with fluorescent ceiling lamps.

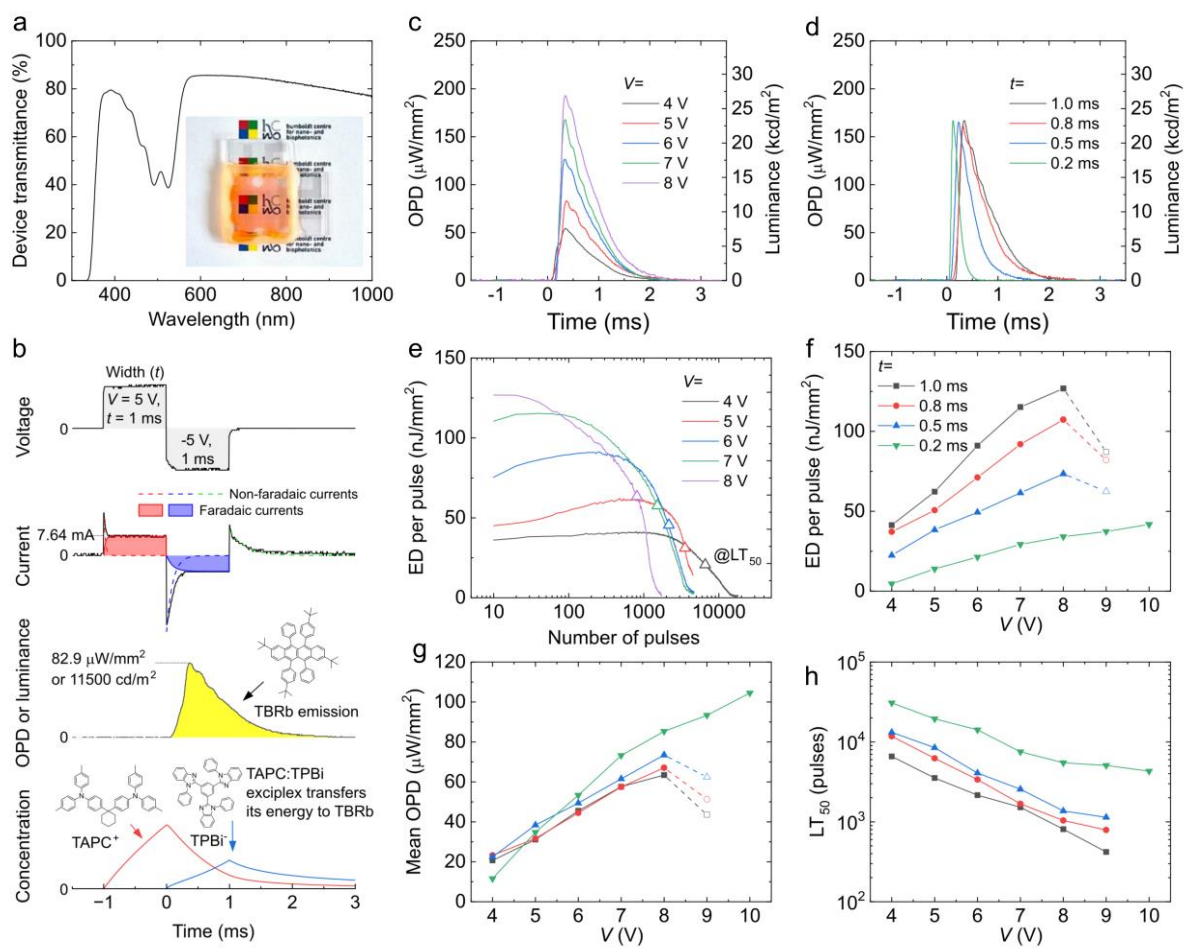

**Fig. 1. Devices operating in pulsed electrochemiluminescence (ECL) mode. a** Transmittance measurement of an ECLD used in this study. Inset shows a photograph of the device on a white paper with logos of our research centre under illumination with fluorescent ceiling lights. The device appears orange due to absorption of the green-blue light by TBRb. **b** Shape of biphasic voltage sequence applied to an ECLD that operates based on an exciplex formation and energy transfer process, and response in current and optical power density (OPD). The bottom graph schematically shows the concentration of TAPC$^+$ and TPBi$^-$ ions near the electrode surface during the biphasic pulse. TAPC$^+$ and TPBi$^-$ ions form exciplexes in the second half of the pulse (time > 0), and subsequently transfer their energy to TBRb emitter molecules. **c** ECL responses to biphasic pulses of different peak voltages (*V*) and with a fixed duration of 1 ms, and **d** for different durations (*t*) at a fixed voltage of 7 V. **e** Evolution of energy density (ED) per pulse as function of the number of applied pulses. Devices were operated with 10 Hz repetition rate and 1.0 ms pulse width. Triangles indicate the estimated LT$_{50}$ lifetime. **f-h** ED, mean OPD, and LT$_{50}$ as functions of peak voltage and pulse width.

Fig. 1b shows the application of a biphasic voltage sequence consisting of ±5 V for 1 ms in each phase, and the resulting current density and optical power density generated by the ECLD. The current consists of faradaic currents arising from electron-transfer processes involving organic molecules and non-faradaic currents due to charging/discharging processes by mobile electrolyte ions near the electrode surfaces. These two components are represented as filled areas (faradaic) and broken lines (non-faradaic); the attribution of each component was made by simulations using an equivalent circuit model (see Ref [27] and Supplementary Fig. S1). A steep rise in ECL intensity is observed during the second half of the biphasic pulse, with the emission peaking at 82.9 µW/mm$^2$ and 11,500 cd/m$^2$ at 0.33 ms into the second half of the pulse. The ECL intensity then gradually decreases over time for about 2 ms. The bottom graph shows a rough estimate of the concentrations of TAPC cations and TPBi anions near the electrode surface based on a simple electrochemical simulation[28]. TAPC cations accumulate during the positive voltage phase, and they react with TPBi anions generated during the negative voltage phase to form exciplexes. The exciplexes then transfer their energy to nearby TBRb molecules, resulting in the onset of light emission. We attribute the small oscillation in emission during the decay in ECL intensity to various long-range coupling processes between TAPC cations and TPBi anions as these oscillations are absent in an ECLD without exciplex materials (see Supplementary Fig. S2). The ECL reaction continues even after the voltage is turned off due to the diffusion of remaining ions. The absolute quantum efficiency of ECL ($\Phi_{ECL}$), defined as the ratio of outcoupled photons to electrons involved in the faradaic current, is estimated to be 0.81%, showing a 1.6-fold enhancement over the 0.52% efficiency from the same device operating under sinusoidal AC operation at 100 Hz[27]. Fig. 1c shows the optical power density response to biphasic pulses of different peak voltages for a fixed width of 1 ms for each phase of the pulse. As the voltage increased from 4 V to 8 V, the peak intensity increased from 54.4 µW/mm$^2$ to 192.0 µW/mm$^2$, corresponding to luminance values of 7.58 kcd/m² and 26.90 kcd/m², respectively. The mean OPD over the duration of the voltage pulse increased from 20.7 µW/mm² to 63.4 µW/mm². The shape of the ECL transient and the response time (time to peak intensity) remained consistent across different voltages. Next, we modified the pulse width ($t$) at a fixed pulse voltage of 7 V (Fig. 1d). As $t$ decreased from 1.0 ms to 0.2 ms, the response time decreased from 0.33 ms to 0.12 ms, also shortening the tail of the ECL transient. The peak intensity remained steady at approximately 167 µW/mm², while the mean intensity increased from 57.6 µW/mm² to 73.2 µW/mm². These observations suggest that the peak ECL intensity is associated with the rate of electrochemical reactions of donor and acceptor materials at the electrodes, which increases at higher voltages. The response time is governed by the speed of charging/discharging processes of mobile electrolyte ions, which becomes faster at shorter pulse widths. In addition, ECL reactions are more efficient at shorter pulse widths as they occur predominantly near electrode surfaces before the ions diffuse away.

Lifetime estimates for ECLDs operated under pulsed operation are shown in Fig. 1e. The energy density (ED) per pulse initially increased with the number of applied voltage pulses due to a redistribution of the concentration of organic substances in the liquid layer under the electric field, and then decreased due to degradation. Higher voltages accelerated degradation due to the increased likelihood of undesired side reactions within the liquid layer. To quantify the operational lifetime, the $LT_{50}$ parameter is defined as the number of pulses applied until the ED reduces to 50% of its maximum; this value is represented by triangles in the Fig. 1e.

Figs. 1f-g summarize the ED, mean OPD, and $LT_{50}$ values at various voltages and pulse widths (also see Supplementary Fig. S3). The ED increases almost linearly with voltage and pulse width, reaching a maximum of 127 nJ/mm² at 8 V and 1 ms pulse width. A further increase in voltage at a pulse width longer than 0.5 ms results in rapid and significant device degradation. In contrast, operation at $t = 0.2$ ms allows for the voltage to be increased to up to 10 V with less degradation, resulting in a maximum mean OPD of 105 µW/mm$^2$. As expected, LT$_{50}$ improves when reducing voltage and pulse width. For the operation conditions corresponding to the maximum ED and maximum mean OPD, LT$_{50}$ was 810 and 4,310 pulses, respectively. The longest operational LT$_{50}$ lifetime of approximately 51 minutes was achieved for $V = 4$ V, $t = 0.2$ ms, and $f = 10$ Hz.

**Modulation of the waveform**

As shown in Fig. 1b, power consumption through the charging/discharging processes is larger in the second voltage phase than in the first, indicating an unbalanced supply of ions and counterions. To mitigate this problem, we further modulated the applied waveform by testing asymmetric biphasic pulses. Fig. 2a shows the current densities and ECL intensities obtained in response to ±5 V pulses with a fixed width of the first phase ($t_1$=1.0 ms) and varying widths of the second phase ($t_2$). Increasing the relative width $t_2/t_1$ extended the faradaic current during the second phase. While the width of the second phase determined the tail length of the ECL transient, it did not affect the ECL peak intensity or response time. The ECL tail length increased up to $t_2/t_1 = 1.25$, suggesting an ionic balance at this pulse ratio. Fig. 2b shows analogous measurements with a fixed voltage during the first phase of the pulse ($V_1 = 5$ V) and varying voltage in the second phase ($V_2$), with each pulse phase lasting 1 ms. Increasing the relative voltage $V_2/V_1$ resulted in higher negative faradaic currents, higher ECL peak intensities, and shorter response times, demonstrating that the voltage during the second phase directly affects the overall ECL dynamics.

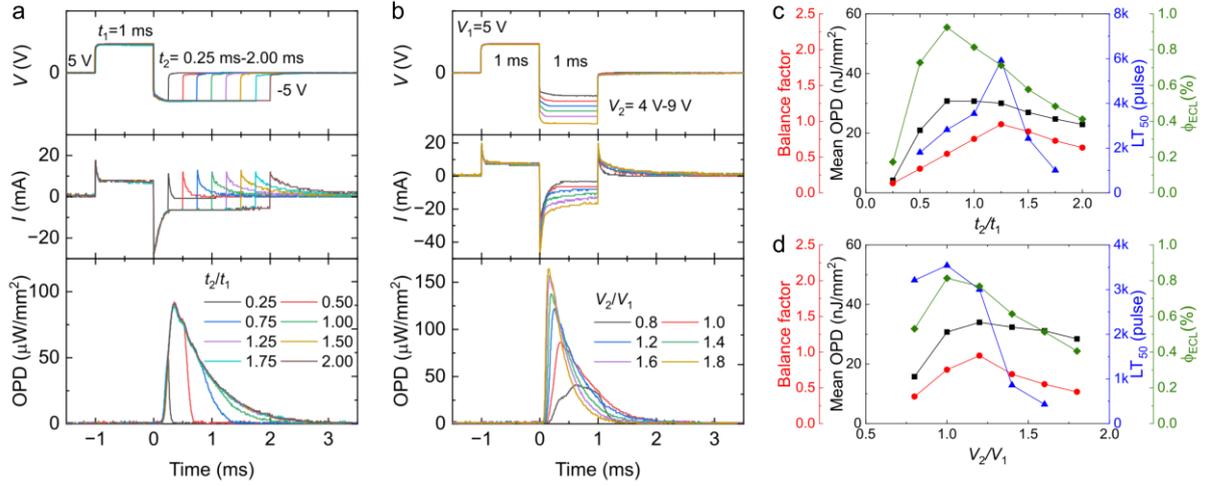

**Fig. 2. Modulation of ECL pulse. a, b** ECL responses to biphasic voltage sequences with varying relative pulse widths ($t_2/t_1$) and voltages ($V_2/V_1$) for the first and second phase of the pulse. The voltage and width during the first phase are fixed to 5 V and 1 ms, respectively. **c,d,** Balance factor for faradaic current, mean OPD, $LT_{50}$, and $\Phi_{ECL}$ as a function of the $t_2/t_1$ and $V_2/V_1$ ratio, respectively.

To quantify the balance between the positive and negative faradaic currents, we introduce a balance factor for the faradaic current, defined as:

$$\text{Balance factor} = \frac{\min(I_{\text{pos}}, I_{\text{neg}})}{\max(I_{\text{pos}}, I_{\text{neg}})}$$

where $I_{\text{pos}}$ and $I_{\text{neg}}$ are the time-integrated faradaic currents during the positive and negative pulses, respectively. A balance factor of 1 represents perfectly balanced current injection, i.e. an ionic balance during device operation. Figs. 2c and 2d summarize the key device characteristics—balance factor, mean OPD, $LT_{50}$, and $\Phi_{ECL}$—as functions of $V_2/V_1$ and $t_2/t_1$, respectively. The ionic balance achieved at $t_2/t_1 = 1.25$ maximizes operational longevity, while an excessive supply of counterions at $t_2/t_1 > 1.25$ significantly reduces the $LT_{50}$ and $\Phi_{ECL}$ values. Both the mean OPD and $\Phi_{ECL}$ values were highest at $t_2/t_1 = 0.75$, indicating again that ECL reactions occurring near the electrode surfaces are more efficient than those occurring due to ion diffusion at a later time point. In contrast, balancing the ion concentrations by increasing the voltage in the second phase did not improve the operational longevity, presumably because the increased likelihood of side reactions at higher voltages outweighs the benefits of ion balancing. As a result, the optimal $LT_{50}$ and $\Phi_{ECL}$ values are achieved at $V_2/V_1 = 1.0$.

Next, we incorporated multiple pulse sequences into a single burst signal to achieve prolonged ECL at high optical power, as is required for the intended application in optogenetics. We introduced a rest period ($p$) of varying duration between sequences to prevent an accumulation of residual ions and to refresh the concentration distribution within the liquid layer. Fig. 3a shows the ECL response to five subsequent voltage pulses, each at ±10 V and with $t_1 = 0.20$ ms and $t_2 = 0.22$ ms to ensure balanced ion concentrations as per our previous optimization (see Supplementary Fig. S4). The rest period was varied from 0 to 2.1 ms. Without rest periods ($p = 0$), ECL peaks appeared during every pulse phase due to the presence of residual ions, and this caused the peak intensity to progressively deteriorate. As the rest period increased in duration, the ECL peaks corresponding to individual biphasic sequences became more distinct, with stable ECL intensity achieved at $t \geq 0.45$ ms (see Supplementary Fig. S5 for the correlation between peak intensity and rest period). Fig. 3b demonstrates that high intensity ECL was maintained for over 0.7 seconds without significant reduction in ECL intensity by incorporating $n = 800$ pulse sequences with a 0.45 ms rest period.

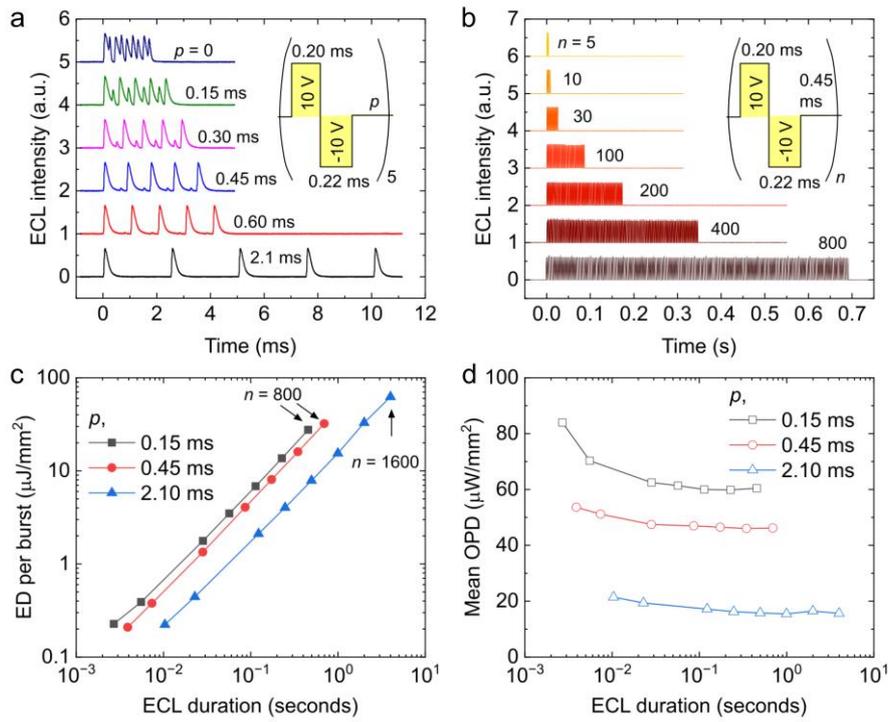

**Fig. 3. Incorporating multiples ECL pulses into a single burst signal. a** Five ECL pulses with varying rest periods ($p$) between biphasic voltage pulses. **b** ECL pulse trains incorporating between $n = 5$ and $n = 800$ biphasic pulses with a rest period of 0.45 ms between pulses. **c** Increase in energy density (ED) and ECL duration with increasing $n$ for three different rest periods. **d** Mean OPD as a function of number of applied voltage pulses for different rest periods.

Figs. 3c and 3d show the total ED and mean OPD for a single burst with increasing number of pulses, incorporating rest periods of 0.15 ms, 0.45 ms, and 2.10 ms. A rest period of 0.45 ms achieved intense and stable light output, yielding a mean OPD of 46.2 µW/mm² over 0.7 seconds and a total ED of 32.1 µJ/mm² at $n=800$ pulses with minimal optical losses introduced with increasing $n$. A longer rest period of 2.10 ms allows for more pulses before device degradation becomes significant, resulting in a mean OPD of 15.6 µW/mm² over 4.0 seconds and a total ED of 62.4 µJ/mm² at $n=1600$ pulses. This configuration is suitable for applications requiring extended illumination periods, albeit with a lower mean OPD compared to the shorter rest periods. Conversely, a shorter rest period of 0.15 ms produces more intense ECL pulses, with an OPD of 60.4 µW/mm² at $n=800$. However, the insufficient removal of radical ions led to a gradual decrease in OPD with increasing number of pulse sequences, ultimately resulting in unstable operation.

**Optogenetic Stimulation of *Drosophila* Larvae**

To show the potential value of ECLDs and the new driving scheme developed here for biomedical and neurophotonics research, we examined the ability of ECLDs to optogenetically activate neurons in in *Drosophila* larvae, which in turn trigger an escape behaviour in the animals. To do this, we placed first or second instar larvae within water droplets of approximately 3 mm diameter located on the ECLDs as illustrated in Fig. 4a. The water droplets allowed us to confine the larvae to the position of an ECLD pixel. We utilized a transgenic *Drosophila* line expressing CsChrimson[29], a light gated cation channel with an excitation spectrum closely aligned with the ECL emission spectrum (Fig. 4b). In this genotype, light pulses activate 'Down and Back' (DnB) interneurons[30], which cause the larvae to bend and roll away in response to perceived threats. Behavioural responses of 10 individual larvae per group were recorded with an inverted microscope. To observe larval behaviour, animals were illuminated with infrared light ($\lambda_{peak}$ = 853 nm, Supplementary Fig. S6), which is outside the excitation wavelength range of CsChrimson, and which can pass through the ECLD without substantial loss. In the absence of ECLD light, the larvae generated peristaltic waves, performed head sweeps, and turned within the water droplet, all characteristics of normal exploratory behaviour.

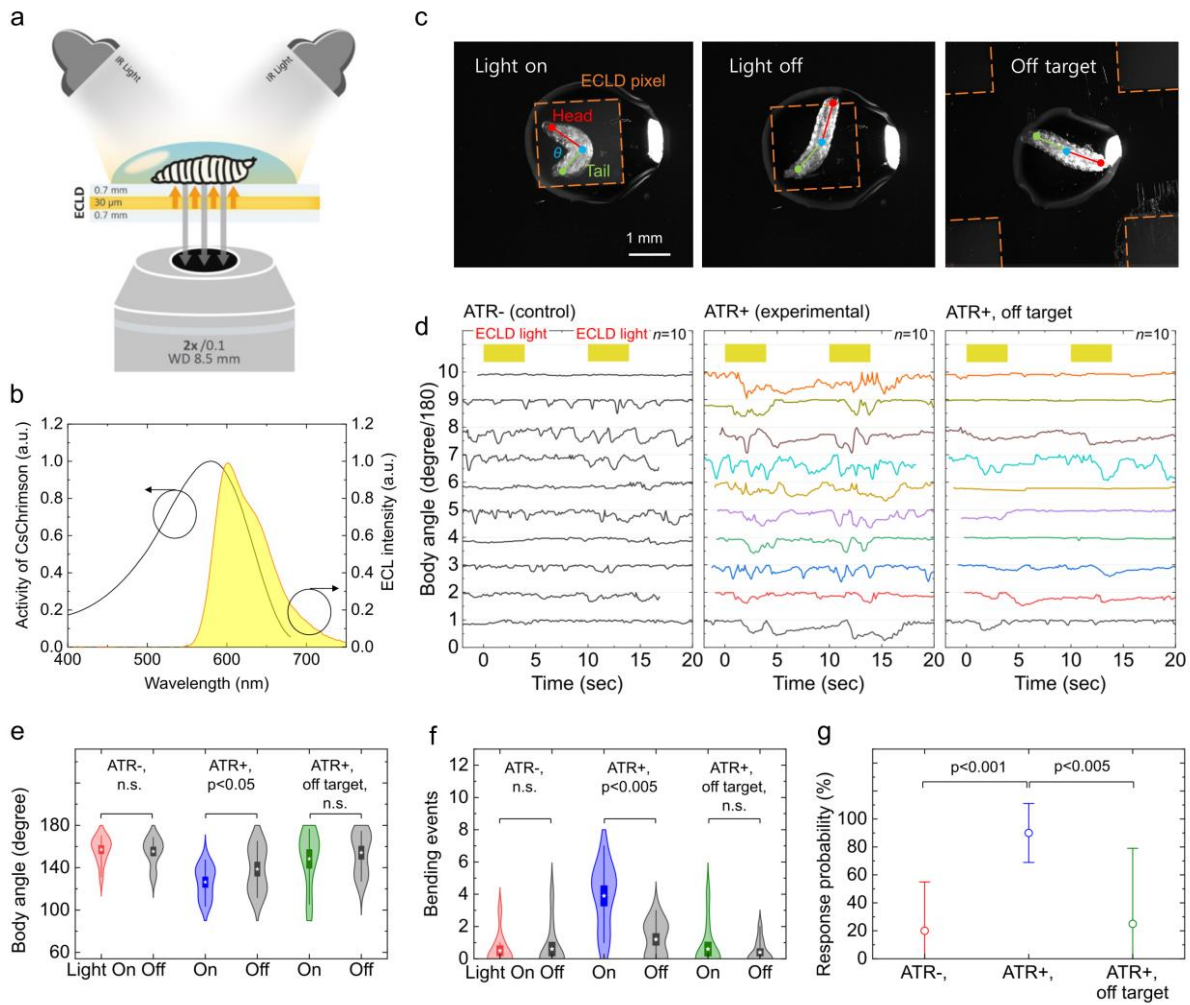

**Fig. 4. Optogenetic manipulation of *Drosophila* larvae behaviour. a** Schematic of the inverted microscope setup. **b** ECL emission spectrum (yellow) and activity spectrum of CsChrimson (black line). **c** Illustration of the three body points (red, blue, green) tracked in larvae to quantify their behavioural response to ECLD illumination, indicated for animal placed on an ECLD pixel with light on (left) and light off (centre), as well as for animal placed away from the active pixel, i.e. off target (right). **d** Body angle over time for 10 individual larvae placed on an ECLD pixel for an ATR-negative control (left) and experimental animals, i.e. fed with ATR supplemented food (centre), as well as for off target ATR+ animal (right). Light stimulation and darkness periods are 4 seconds and 6 seconds, respectively. **e,f** Statistical analyses of body angle and bend events for larvae in experimental and control groups as well as off target group during light "on" and "off" phases. **g** Response probability in control and experimental groups, highlighting the likelihood of bending in response to ECL light stimulation.

To test the behavioural response upon optogenetic stimulation, we tracked the angle between head and tail of the larvae (referred to as "body angle" $\theta$ in Fig. 4c) during 25-second-long trials which consisted of alternating 4-second light "on" and 6-second light "off" phases. During the "on" phases, the ECLD operated at a mean OPD ranging from 1.84 µW/mm$^2$ (±4V, $n$ = 1600, $p$ = 2.10 ms), which is near the light sensitivity threshold of CsChrimson in *Drosophila* larvae (2.0 µW/mm$^2$ at 90% response probability, as estimated by Murawski *et al.*[31]), to 15.6 µW/mm$^2$ (±10V).

Functional expression of CsChrimson requires the presence of all-trans-retinal (ATR) as co-factor. When food was not supplemented with ATR (ATR- control group), ECLD light pulses did not induce significant changes in body angle (Fig. 4d). In contrast, larvae in the experimental group displayed a marked decrease in body angle during the "on" phases compared to the "off" phases. The average body angles during "on" and "off" phases were measured at 126° and 139°, respectively (Fig. 3e, $p<0.05$). Note that the onset of the larvae response was often delayed, and the reduced body angle persisted after the ECLD was turned off, which was included in the "off" phase statistics, indicating that effects may be even stronger than reported here.

Trials conducted with larvae positioned approximately 2.1 mm to the side of the active pixel ("off-target" trials, Fig. 4c) did not show significant changes in the body angle, with average body angles of 148° and 153° during the "on" and "off" phases, respectively. These results confirm that ECLDs can elicit an optogenetic response with high spatial resolution, while also corroborating that behavioural responses are due to light stimulation rather than by any electrical, chemical, or thermal effect from device operation.

Further analysis involved counting the number of "bend" events, where the body angle dropped below 90 degrees for at least 100 ms during the "on" phase. Larvae in the experimental group showed a highly increased number of bend events during the "on" phases compared to the "off" phases (Fig. 3f, *p*<0.005). In contrast, neither the off-target group nor ATR- control group showed any significant differences between light "on" and "off" phases.

To quantify the likelihood of larvae reacting with a bend event to ECLD light, we calculated the response probability of each larva as the ratio of stimulations evoking a behavioural response, to the total number of stimulations (see Methods for response criteria). The ATR-control group set a baseline probability of 19% due to spontaneous movement. In contrast, larvae in the experimental group had a 90% probability of reacting to the ECL pulses (Fig. 3g, *p*<0.001). The response probability of the off-target group was not significantly different from the baseline of the control group.

**Discussion**

Due to insufficient light output and low operational stability, ECLDs have so far faced challenges in lighting applications. This study demonstrated a significant improvement in ECL intensity and lifetime through a thorough optimization of the pulsed operation of ECLDs. A robust electrochemical operating mechanism based on exciplex formation on the molecules TAPC and TPBi facilitated the application of voltage up to 10 V in the pulsed operation. These improvements were applied to highlight the potential of ECLDs as a light source capable of delivering stable, high-power light emission for biomedical research, in particular for optogenetics experiments.

Intense ECL pulses were generated through the use of an exciplex-based reaction pathway followed by energy-transfer to a terminal emitter and applied biphasic voltage sequences to the device that consisted of pulses with opposing voltage lasting of order 1 ms each. We observed a rapid increase in ECL intensity during the second half of each biphasic sequence arising from a reaction of counter ions with ions accumulated during the initial phase of the voltage pulse.

The subsequent gradual decrease in intensity over time was attributed to slow ion diffusion. The quantum efficiency of the pulse operation was estimated to be 0.81%, which is 1.6-fold higher than the 0.52% efficiency previously reported for the same emitter system but under standard AC operation[27]. We measured an exceptionally high mean OPD of 63.4 µW/mm² with a peak OPD of 192 µW/mm² in an ECL pulse for a biphasic voltage sequence of ±8 V and 1 ms duration, with an ED of 127 nJ/mm² per pulse. Reducing the pulse width allowed for a further increase in voltage, achieving an even higher mean OPD of 104 µW/mm² at ±10 V and 0.2 ms pulse duration; 4310 emission pulses were delivered in this configuration until 50% degradation.

A modest extension of the width of the second phase of the biphasic voltage sequence was found to effectively increase the operational lifetime through balancing ion concentrations. Increasing the voltage during the second phase of the biphasic pulse enhanced the mean OPD and shortened the response time but was not effective in increasing the longevity due to enhanced side reactions.

Finally, we incorporated multiple optimized voltage sequences into a single burst signal to extend the duration of ECL of high optical power. Inserting a rest period longer than 0.45 ms refreshed the device between sequences, allowing for more than a thousand ECL pulses without noticeable device degradation. As a result, a single burst signal maintained stable ECL pulses at an optical power of 60.4 µW/mm² for 0.46 seconds using a 0.45 ms rest period, and at an optical power of 15.6 µW/mm² for 4.0 seconds using a 2.10 ms rest period.

To demonstrate their potential for biomedical applications, we used ECLDs to optogenetically stimulate and observe *Drosophila* larvae. The light output of the ECLD operating by optimized pulse sequences effectively and reliably induced escape behaviours in larvae expressing CsChrimson in an ensemble of neurons. When exposed to alternating 4-second light "on" phases and 6-second light "off" phases, the larvae exhibited a significantly reduced body angle due to a body-bending behaviour during the "on" phases (126°) compared to the "off" phases (139°). This was in line with the first stage of rolling behaviour observed by Burgos *et al*.[30] Larvae in the experimental group demonstrated a striking 90% response rate to optogenetic stimulation, significantly higher than the 19% baseline observed in the control group. In addition, the larvae in the experimental group showed negligible response to light when they were positioned away from the ECLD pixels, demonstrating the high spatial resolution of optogenetic triggering using ECLDs.

Optogenetics is a widely used tool for studying the nervous system across multiple model organisms and fields of biomedical sciences. However, traditional light sources make imaging during optogenetics experiments challenging as the opaque nature of most light sources requires careful placement of the microscope or camera away from the light source or use of multiple light paths and/or dichroic mirrors and filters. Our semi-transparent ECLDs open up new opportunities by providing the ability to image through the light source, placing the microscope or camera directly in line with the animal or tissue of interest. In addition, the light source can be placed directly in contact with the target animal or tissue if needed, thus opening a broad range of experimental configurations for optogenetics, biomedical research and neurophotonics in general. Overall, these experiments show the potential for ECLDs to not only be an effective illumination tool in optogenetics but to provide a platform for a range of experiments not possible with conventional LEDs.

## Materials and methods

### Materials

TAPC, TCTA, and TBRb were purchased from Luminescence Technology Corp. and used as received. Anhydrous toluene, anhydrous acetonitrile, and TBAPF$_6$ were purchased from Merck KGaA. ITO-coated glass substrates were purchased from Xinyan Technology Ltd. The encapsulation glue (3035BT) was purchased from Threebond International.

### Device fabrication

To manufacture an ECLD, a pair of ITO-coated rectangular glass substrates with a size of 24 mm by 15 mm were cleaned by immersion in a detergent solution (2% Hellmanex® III in Milli-Q water) and subsequently in isopropyl alcohol. The cleaned substrates were dried on a hot plate. The surfaces of ITO were treated with UV-ozone for 15 minutes. The glass substrates were bonded together with UV-curable NOA68 resin (Norland Products) at an angle of 90°, where the ITO surfaces of the two substrates faced each other and intersected with an overlapping area of 4 mm$^2$. Polystyrene microbeads were mixed in the resin to keep the gap between glass substrates at 30 μm. The bonded substrates were then transferred to a nitrogen-atmosphere glove box. 6.5 μL of ECL solution was pipetted into gap between substrates for capillary filling. The edges of the liquid-filled area were sealed by 3035BT (Threebond International) resin and then cured under UV light. The details of this procedure are given in Ref [26].

### Device characterization

To characterize the electroluminescence properties of the devices, we employed the setup previously detailed by Moon et al.[26] In brief, this setup included an arbitrary waveform generator (33220A, Agilent Technologies) that allowed to drive the devices with biphasic rectangular pulses, varying in frequency, voltage, and pulse width for both the positive and negative voltage parts as described in the main text. The waveform was programmed via the SCPI interface using a custom Python software. During the pulsing, the device current was measured by recording the voltage drop across a 1 Ohm shunt resistor in series with the ECLD, using an oscilloscope (Rigol DS1000Z). The brightness of the devices was monitored using a calibrated silicon photodiode (PDA100A2, Thorlabs), which was connected to the second channel of the same oscilloscope for temporal readout. The photodiode was positioned 168 mm from the device on one side, while emissions from the opposite side were disregarded in the analysis. Spectral data were obtained using a fiber-coupled spectrometer (Ocean HDX, Ocean Insight) and used to convert the brightness measurements into absolute units, assuming a Lambertian emission profile. For lifetime measurements, device brightness was recorded at one-second intervals over extended durations.

### Drosophila culture

Transgenic *Drosophila* flies with 412-GAL4 (a.k.a DnB-GAL4) stably combined with 20XUAS-IVS-CsChrimson coupled to a mVenus fluorophore were used for all experiments. 412-GAL4 expresses in DnB interneurons, which trigger larval escape behaviours when activated (29). Animals were raised in vials on standard cornmeal-based food at 20°C on approximately 12:12 light-dark cycles. Two weeks before the optogenetic experiments, eggs were transferred onto fresh vials with 5 ml media and 0.725 mM all-trans retinal (Sigma Aldrich, USA) final concentration. Vials were kept in the dark at 20 °C and away from heat. ATR (-) control groups for the optical stimulation were 412-GAL4; 20XUAS-IVS-CsChrimson-mVenus larvae, kept under the same conditions as the optogenetic group, but without ATR supplemented food.

*Drosophila* imaging

Imaging was performed on a standard inverted microscope (Tie2, Nikon, Japan) using a Plan Apo 2x objective with a numerical aperture of 0.1 and a working distance of 8.5 mm (Nikon, Japan). Two infrared light sources (IR49S, Andoer, China) were installed above the microscope stage to illuminate the larvae. Videos with 2048 by 2048-pixel resolution, 20 frames per second and 16 bit depth were recorded with an sCMOS camera (Orca Flash 4.0, Hamamatsu, Japan) controlled by the NIS Elements Suite (Nikon, Japan). First or second instar larvae were picked from the vials under low ambient light, briefly washed in water and transferred into a drop of water on the surface of the ECLD above an active pixel. The larvae were habituated for two to three minutes without ECLD light before the experiment was started. After the stimulation of the larva on top of a pixel was completed, the same larva was moved to an off-pixel area with a fresh drop of water and the experiment was run again after a habituation period. -ATR control larvae were only imaged on pixels instead of both on and off pixels. Larvae were only partially submerged for 5-10 min in any given experiment and had periodic access to air through posterior spiracles.

**Data Analysis and Statistics**

Videos were analyzed using the manual tracking plugin of ImgeJ2 (Version 2.14.0). Further analyses were performed with custom scripts in Matlab 2023a (MathWorks) and Origin 2024 (Origin Lab). The head, tail and center of the larvae were tracked and the body bending angle was calculated using the following equation:

$$\cos\theta = \frac{\overrightarrow{\text{head}} \times \overrightarrow{\text{tail}}}{|\overrightarrow{\text{head}}| \times |\overrightarrow{\text{tail}}|}$$

All statistical testing was done using two samples t-tests with an alpha of 5%. The overall body during light on and off phases were compared within each experimental group (ATR-, ATR+, ATR+, off target). To analyze specific behavioral responses to light stimulation, curl events per light on and off phase per trial were counted. A curl event was automatically detected by thresholding during data analysis when the body angle $\theta$ was below 90° for a minimum of 100 ms and compared within each experimental group. To further correlate a curl event to a light on phase, a response probability was calculated based on the ratio of responses to stimulations during light on phases and compared between groups.

## Acknowledgements

This work was financially supported by the Alexander von Humboldt Foundation (Humboldt-Professorship to M.C.G.). C.-K.M. acknowledges funding from the European Commission through a Marie Skłodowska Curie individual fellowship (101029807). J.F.B. acknowledges funding from Beverly and Frank MacInnis via the University of St Andrews. R. S. acknowledges funding through the Global PhD programme at the University of St Andrews and the RS MacDonald Trust.

## Author contributions

C.-K.M. conceived the project, manufactured ECLDs, and characterized them. M. K. and R. S. organized the *Drosophila* experiment and the microscopy. J.F.B. developed the software for device characterization and synchronized the device operation with microscope imaging. S. R. P. supervised the *Drosophila* rearing and experimentation and assisted with writing the manuscript. M.C.G. supervised the research. C.-K.M. and M.C.G. wrote the manuscript with input from all authors.

## Data availability

All primary data for all figures and Supplementary figures are available from the corresponding author upon request.

## Competing interests

The authors declare no competing interests.

## References

1  Kang, C.-M. & Lee, H. Recent progress of organic light-emitting diode microdisplays for augmented reality/virtual reality applications. *J. Inf. D.* **23**, 19-32 (2022).
2  Song, J., Lee, H., Jeong, E. G., Choi, K. C. & Yoo, S. Organic light-emitting diodes: pushing toward the limits and beyond. *Adv. Mater.* **32**, 1907539 (2020).
3  Murawski, C. & Gather, M. C. Emerging biomedical applications of organic light-emitting diodes. *Adv. Opt. Mater.* **9**, 2100269 (2021).
4  Tokel-Takvoryan, N. E., Hemingway, R. E. & Bard, A. J. Electrogenerated chemiluminescence. XIII. Electrochemical and electrogenerated chemiluminescence studies of ruthenium chelates. *J. Am. Chem. Soc.* **95**, 6582-6589 (1973).
5  Cho, K. G. *et al.* Light-Emitting Devices Based on Electrochemiluminescence Gels. *Adv. Funct. Mater.* **30** (2020).
6  Blackburn, G. F. *et al.* Electrochemiluminescence detection for development of immunoassays and DNA probe assays for clinical diagnostics. *Clin. Chem.* **37**, 1534-1539 (1991).
7  Zhan, W. & Bard, A. J. Electrogenerated chemiluminescence. 83. Immunoassay of human C-reactive protein by using Ru(bpy)$_3^{2+}$-encapsulated liposomes as labels. *Anal. Chem.* **79**, 459-463 (2007).
8  Marquette, C. A. & Blum, L. J. Electro-chemiluminescent biosensing. *Anal. Bioanal. Chem.* **390**, 155-168 (2008).

9   Gao, W., Saqib, M., Qi, L., Zhang, W. & Xu, G. Recent advances in electrochemiluminescence devices for point-of-care testing. *Curr. Opin. Electrochem.* **3**, 4-10 (2017).
10  Ma, C., Cao, Y., Gou, X. & Zhu, J.-J. Recent progress in electrochemiluminescence sensing and imaging. *Anal. Chem.* **92**, 431-454 (2019).
11  Bansal, A., Yang, F., Xi, T., Zhang, Y. & Ho, J. S. In vivo wireless photonic photodynamic therapy. *PNAS* **115**, 1469-1474 (2018).
12  Kim, D. *et al.* Ultraflexible organic light-emitting diodes for optogenetic nerve stimulation. *PNAS* **117**, 21138-21146 (2020).
13  Keppeler, D. *et al.* Multichannel optogenetic stimulation of the auditory pathway using microfabricated LED cochlear implants in rodents. *Sci. Trans. Med.* **12**, eabb8086 (2020).
14  Kathe, C. *et al.* Wireless closed-loop optogenetics across the entire dorsoventral spinal cord in mice. *Nat. Biotech.* **40**, 198-208 (2022).
15  Taal, A. J. *et al.* Optogenetic stimulation probes with single-neuron resolution based on organic LEDs monolithically integrated on CMOS. *Nat. Electron.* **6**, 669-679 (2023).
16  Itoh, N. Electrochemical light-emitting gel made by using an ionic liquid as the electrolyte. *J. Electrochem. Soc.* **156**, J37 (2008).
17  Moon, H. C., Lodge, T. P. & Frisbie, C. D. Solution-processable electrochemiluminescent ion gels for flexible, low-voltage, emissive displays on plastic. *J. Am. Chem. Soc.* **136**, 3705-3712 (2014).
18  Kwon, D.-K. & Myoung, J.-M. Ion gel-based flexible electrochemiluminescence full-color display with improved sky-blue emission using a mixed-metal chelate system. *Chem. Eng. J.* **379**, 122347 (2020).
19  Kim, S. *et al.* All-Printed Electrically Driven Lighting via Electrochemiluminescence. *Adv. Mater. Technol.* 2302190 (2024).
20  Laser, D. & Bard, A. J. Electrogenerated Chemiluminescence: XXIII. On the Operation and Lifetime of ECL Devices. *J. Electrochem. Soc.* **122**, 632 (1975).
21  Nobeshima, T., Morimoto, T., Nakamura, K. & Kobayashi, N. Advantage of an AC-driven electrochemiluminescent cell containing a Ru(bpy)$_3^{2+}$ complex for quick response and high efficiency. *J. Mater. Chem.* **20** (2010).
22  Nobeshima, T., Nakamura, K. & Kobayashi, N. Reaction Mechanism and Improved Performance of Solution-Based Electrochemiluminescence Cell Driven by Alternating Current. *Jap. J. Appl. Phys.* **52** (2013).
23  Oh, H., Kim, Y. M., Jeong, U. & Moon, H. C. Balancing the Concentrations of Redox Species to Improve Electrochemiluminescence by Tailoring the Symmetry of the AC Voltage. *ChemElectroChem* **5**, 2836-2841 (2018).
24  Soulsby, L. C. *et al.* Colour tuning and enhancement of gel-based electrochemiluminescence devices utilising Ru (ii) and Ir (iii) complexes. *Chem. Comm.* **55**, 11474-11477 (2019).
25  Ko, E.-S. *et al.* Pulsed Driving Methods for Enhancing the Stability of Electrochemiluminescence Devices. *ACS Photonics* **5**, 3723-3730 (2018).
26  Moon, C. K., Butscher, J. F. & Gather, M. C. An Exciplex-Based Light-Emission Pathway for Solution-State Electrochemiluminescent Devices. *Adv. Mater.* **35**, 2302544 (2023).
27  Moon, C. K. & Gather, M. C. Absolute quantum efficiency measurements of electrochemiluminescent devices through electrical impedance spectroscopy. *Adv. Opt. Mater.* **12**, 2401253 (2024).
28  Lee, J. I. *et al.* Dynamic Interplay between Transport and Reaction Kinetics of Luminophores on the Operation of AC-Driven Electrochemiluminescence Devices. *ACS Appl. Mater. Interfaces* **10**, 41562-41569 (2018).
29   Klapoetke, N. C. *et al.* Independent optical excitation of distinct neural populations. *Nat. Methods* **11**, 338-346 (2014).
30   Burgos, A. *et al.* Nociceptive interneurons control modular motor pathways to promote escape behavior in Drosophila. *Elife* **7**, e26016 (2018).
31   Murawski, C., Pulver, S. R. & Gather, M. C. Segment-specific optogenetic stimulation in Drosophila melanogaster with linear arrays of organic light-emitting diodes. *Nat. Comm.* **11**, 6248 (2020).

**Supplementary Information**

# High-power pulsed electrochemiluminescence for optogenetic manipulation of Drosophila larval behaviour

Chang-Ki Moon[1,2], Matthias König[1,2], Ranjini Sircar[1,3], Julian F. Butscher[1,2], Stefan R. Pulver[3], Malte C. Gather[1,2]

[1]Humboldt Centre for Nano- and Biophotonics, Institute for Light and Matter, Department of Chemistry, University of Cologne, Greinstr. 4-6, 50939 Cologne, Germany

[2]Centre of Biophotonics, SUPA, School of Physics and Astronomy, University of St Andrews, North Haugh, St Andrews KY16 9SS, United Kingdom

[3]School of Psychology and Neuroscience, University of St Andrews, St Mary's St Mary's Quad, South St, St Andrews KY16 9JP, United Kingdom

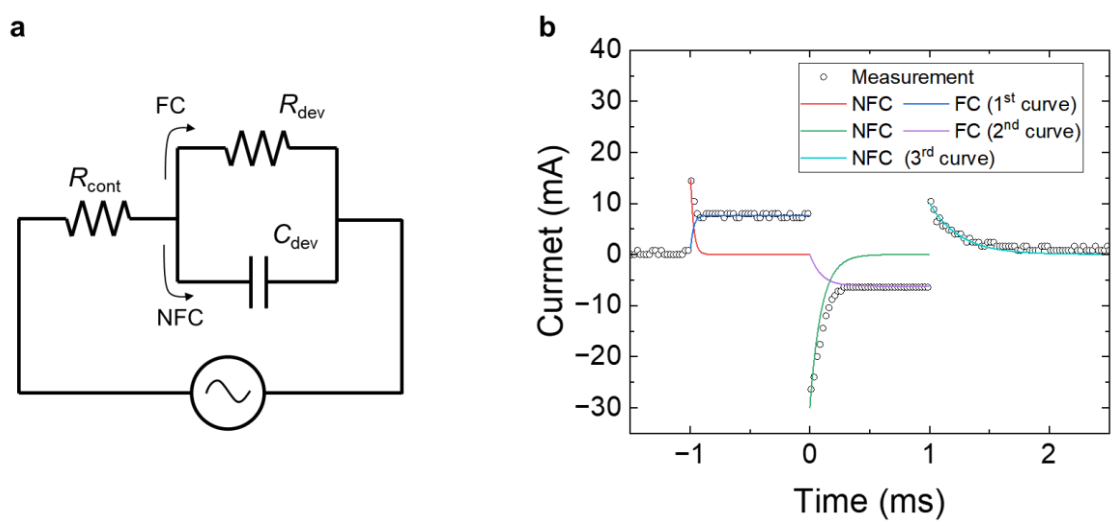

**Supplementary Fig. 1. Modeling of the faradaic current (FC) and non-faradaic current (NFC). a** Equivalent circuit of ECLD. $R_C$, $R_{dev}$ and $C_{dev}$ represent the contact resistance, device resistance, and device capacitance, respectively. **b** Simulated FCs and NFCs for an ECLD operated by a biphasic voltage sequence with a height of 5 V and a width of 1 ms. The time constants are $2.5\times10^{-5}$ s, $9.6\times10^{-5}$ s, and $2.5\times10^{-4}$ s, respectively.

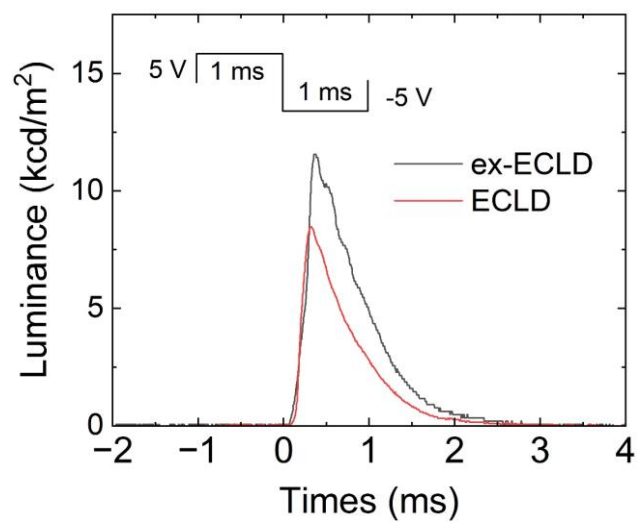

**Supplementary Fig. 2**. **Transient electrochemiluminescence (ECL) response of devices using exciplex materials in addition to emitter (black curve) and solely emitter (red curve).** The biphasic voltage sequence consists of +5 V for 1 ms followed by -5 V for another 1 ms. A decay with multiple small peaks over time is observed only from the device using exciplex materials, indicating that the multiple peaks result from long-range coupling of exciplexes.

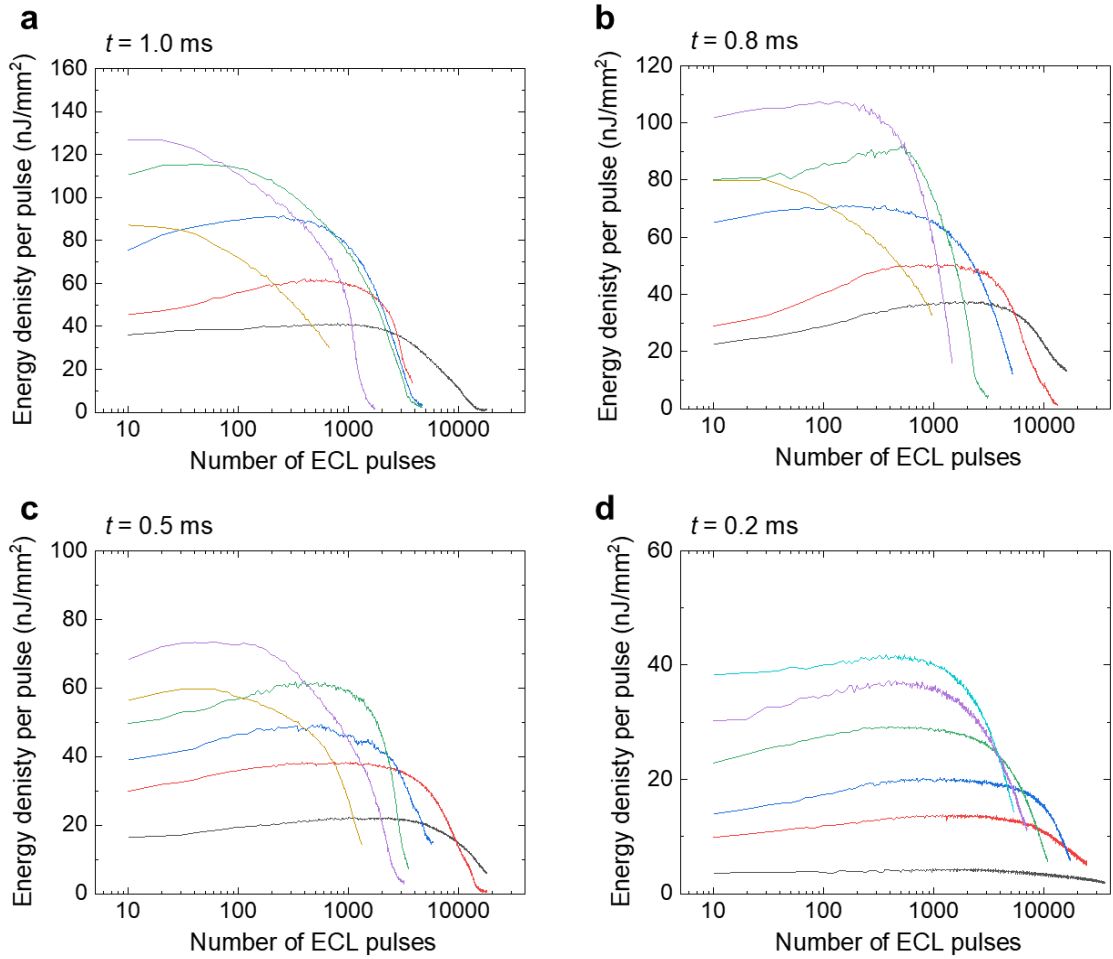

**Supplementary Fig. 3. Estimate of lifetime under pulsed operation. a-d** Energy density (ED) of ECL pulse versus number of applied biphasic voltage sequences for various voltage heights (*V*) ranging from 4 V to 10 V and various widths (*t*) ranging from 1.0 ms to 0.2 ms. All devices operated at a frequency of 10 Hz.

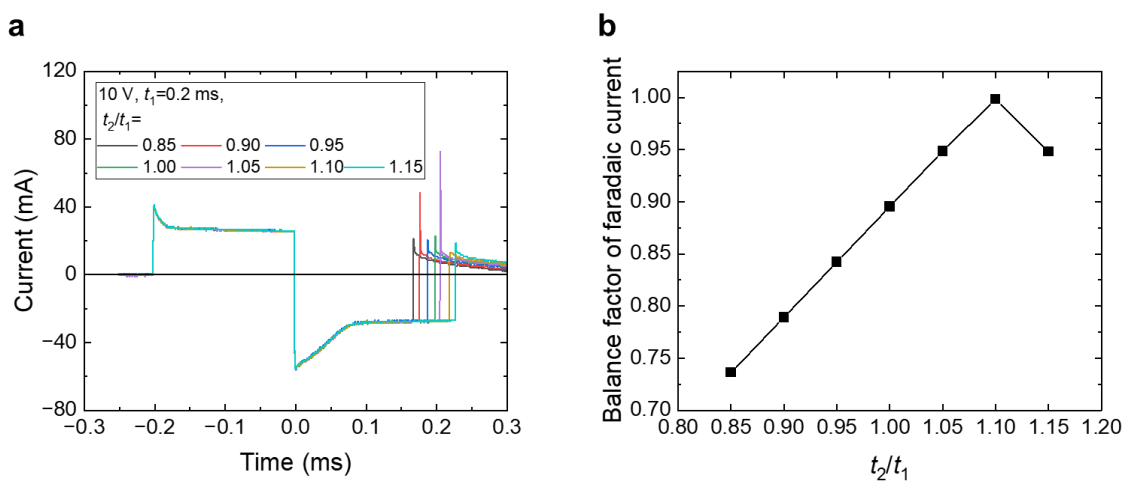

**Supplementary Fig. 4. Balance factor estimation. a** Device current upon application of +10 V for 0.20 ms followed by -10 V for a time ranging from 0.17 ms to 0.23 ms. **b** Balance factor of faradaic current as a function of relative widths of the positive and negative voltage phases ($t_2/t_1$). A balanced injection of charges is measured at $t_2/t_1=1.10$.

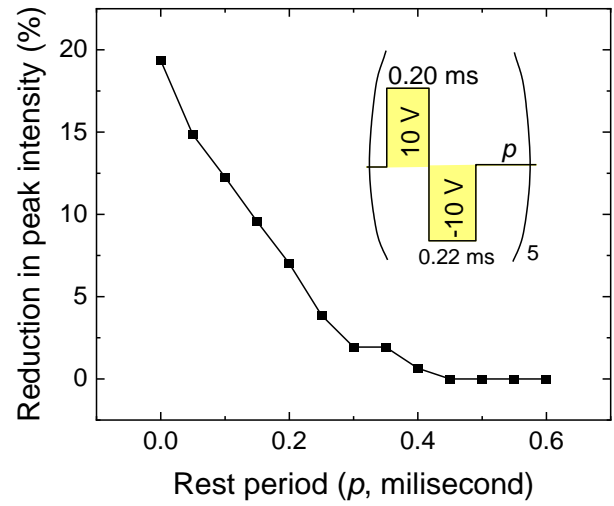

**Supplementary Fig. 5. Reduction in ECL peak intensity in 5th pulse depending on the rest period (*p*) between biphasic voltage sequences.** The voltage sequence consists of 10 V for 0.20 ms for the first phase and -10 V for 0.22 ms for the second phase.

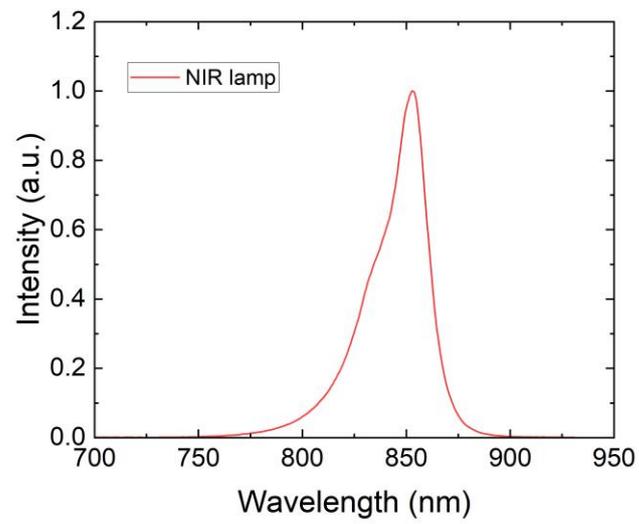

**Supplementary Fig. 6. Infrared lamp spectrum used for the *Drosophila Larvae* experiments.**